\documentclass[fleqn,12pt,twoside]{article}
\usepackage{espcrc1}
\usepackage{graphicx}
\usepackage{epsfig}
\usepackage[figuresright]{rotating}

\newcommand{\AmS}{{\protect\the\textfont2
  A\kern-.1667em\lower.5ex\hbox{M}\kern-.125emS}}
\title{Results on Identified Hadrons from the PHENIX Experiment at RHIC}
\author{T. Chujo\address[BNL]{Brookhaven National Laboratory, Upton, NY 11973-5000, USA} 
for the PHENIX Collaboration\thanks{for the full PHENIX Collaboration author
list and acknowledgements, see Appendix ``Collaborations'' of this volume.}
}
       
\begin{document}
\maketitle

\begin{abstract}
Recent results on identified hadrons from the PHENIX experiment in Au+Au 
collisions at mid-rapidity at $\sqrt{s_{NN}}$ = 200 GeV are presented. 
The centrality dependence of transverse momentum distributions and particle 
ratios for identified charged hadrons are studied. The transverse
flow velocity and freeze-out temperature are extracted from $p_{T}$ 
spectra within the framework of a hydrodynamic collective flow model. 
Two-particle HBT correlations for charged pions are measured in different 
centrality selections for a broad range of transverse momentum of the pair. 
Results on elliptic flow measurements with respect to the reaction 
plane for identified particles are also presented. 
\end{abstract}

\section{INTRODUCTION}
The physics motivation of the ultra-relativistic heavy-ion program at 
the Relativistic Heavy Ion Collider (RHIC) is to study nuclear matter 
at extremely high temperature and energy density with the hope to reach 
a new form of matter called the quark gluon plasma (QGP). Among the various 
probes of the QGP state, hadrons carry important information about the 
collision dynamics along with the spatial and temporal evolution of the 
system from the early stage of the collisions to the final state interactions.

For studies of hadron physics at RHIC, the PHENIX experiment~\cite{NIM} 
demonstrates good capability for particle identification (PID) for both 
charged hadrons ($\pi^{\pm}$, $K^{\pm}$, $p$, $\overline{p}$, $d$ and 
$\overline{d}$) and neutral pions over a broad momentum range. The charged 
hadrons can be identified with time-of-flight measurements in two different 
detectors: (1) a high resolution Time-of-Flight wall (TOF) and, 
(2) an electro-magnetic calorimeter (EMC), in conjunction with the tracking 
system in the PHENIX central arm spectrometers and the beam counter, which 
provides the start timing and the event vertex determination. The tracking 
system in the central arm consists of drift chambers (DC), three layers of 
pad chambers (PC), and time expansion chambers (TEC). The PHENIX central arms 
cover $|\eta|<0.35$ in pseudo-rapidity, and cover $\pi/4$ with the TOF and 
$3\pi/4$ by EMC in azimuth. The $\pi/K$ and $K/p$ separation can be achieved 
up to 2 and 4 GeV/$c$ in $p_{T}$, respectively, using the TOF detector, which 
has a 120 ps timing resolution. The EMC offers a larger azimuthal coverage 
for PID, but suffers from a timing resolution of only 500 ps. Neutral pions 
are identified with the EMC via the $\pi^0 \rightarrow \gamma\gamma$ decay 
channel up to 10 GeV/$c$ in $p_{T}$ using the full statistics taken during 
Run II at RHIC in Au+Au collisions~\cite{David}.  

\section{IDENTIFIED SINGLE PARTICLE SPECTRA}

\begin{figure}[t]
\begin{center}
\includegraphics[width=14cm]{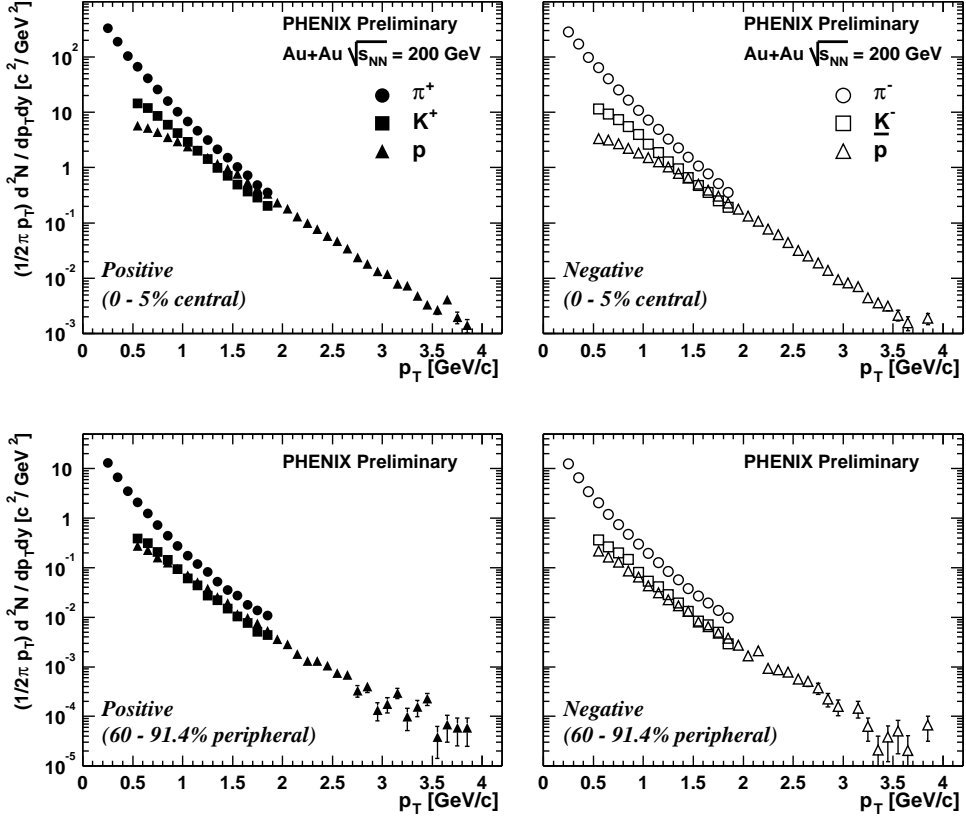}
\caption{Transverse momentum distributions for pions (circles), 
kaons (squares) and $p$, $\overline{p}$ (triangles) in the 0--5\% 
most central events (upper panels) and 60--91.4\% most
peripheral events (lower panels) at $\sqrt{s_{NN}}$ = 200 GeV 
in Au+Au collisions. The left panels show positive particles 
and the right panels show negative particles . The error bars are 
statistical only.}
\label{fig:Spectra_cent}
\end{center}
\end{figure}

\begin{figure}[h]
\begin{center}
\includegraphics[width=14cm]{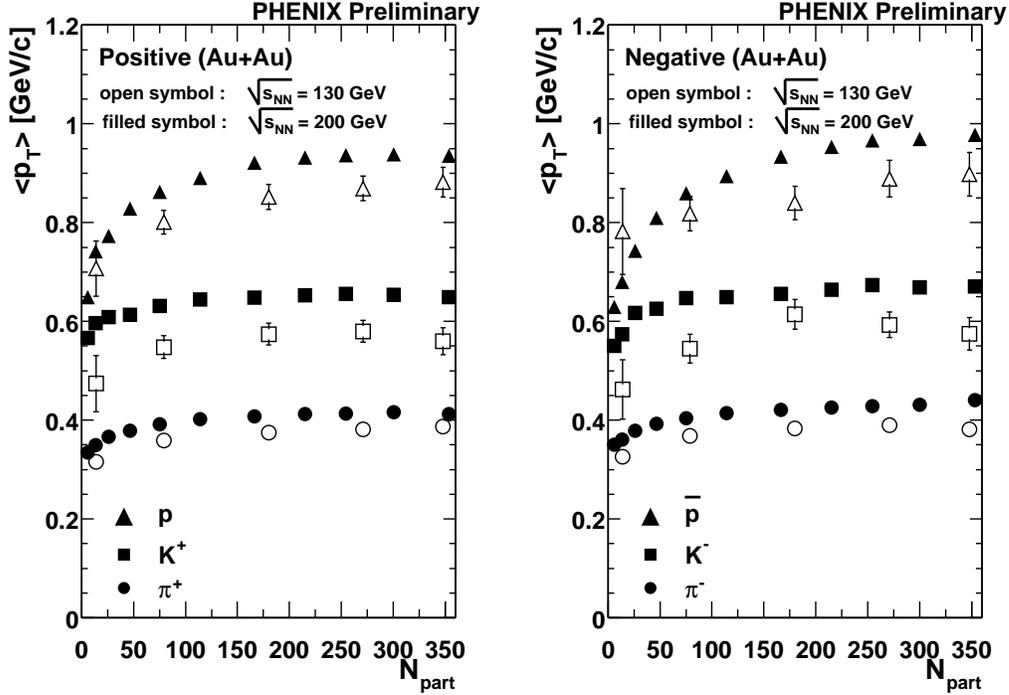}
\caption{Mean transverse momentum for identified charged
hadrons as a function of $N_{part}$ for protons and anti-protons 
(triangles), kaons (squares) and pions (circles). The left panel
shows positive particles and right panel shows negative particles. 
The open symbols indicate the data for 130 GeV~\cite{PPG006} while the 
filled symbols indicate the data for 200 GeV in Au+Au collisions.}
\label{fig:meanPt}
\end{center}
\end{figure}

We have measured the transverse momentum distributions for $\pi^{\pm}$, 
$K^{\pm}$, $p$ and $\overline{p}$ at mid-rapidity in Au+Au collisions 
at $\sqrt{s_{NN}}$ = 200 GeV over a broad momentum range over various 
centrality selections. To identify the charged particles, the high 
resolution TOF counter is used in this analysis. We use about 4 million 
minimum bias events. The data are classified into 11 centrality bins
expressed in percent of the total inelastic cross section. 
The spectra for each particle species are corrected for geometrical 
acceptance, decay in flight, multiple scattering, and tracking efficiency 
using a single particle Monte Carlo simulation. A multiplicity-dependent 
track reconstruction efficiency is also determined and applied by embedding 
simulated tracks into real events.

The upper two panels in Figure~\ref{fig:Spectra_cent} show the $p_T$ 
distributions for identified hadrons in the most central events (0--5\%) 
in 200 GeV Au+Au collisions for positive (left) and negative (right)
particles. The lower two panels show the most peripheral events (60--91.4\%).
In each panel, the data are presented up to 1.8 GeV/$c$ for charged 
pions and kaons, and 3.8 GeV/$c$ for $p$ and $\overline{p}$. In the 
low $p_T$ region of the most central events, the data indicate that the 
inverse slope increases with the particle mass. Also, the shape of the $p$ 
and $\overline{p}$  spectra have a shoulder-arm shape while the pion spectra 
have a concave shape. On the other hand, in the most peripheral events, the 
spectra are almost parallel to each other. This mass dependence of the slopes 
and shapes of the spectra for protons in the central events can 
be explained by a radial flow picture. At around 2.0 GeV/$c$  
in central events, the proton yield becomes comparable to the pion yield.
A similar behavior is also seen for negatively charged particles. 

In order to quantify the observed mass dependence of the slopes,
the mean transverse momenta $\langle p_{T} \rangle$ are extracted 
for 11 centrality selections and for each particle species.
The $p_T$ spectra are extrapolated to below and above the measured range
using a power law function for pions, an $m_{T}$ exponential function for 
kaons, and a Boltzmann function for $p$ and $\overline{p}$. 
In Figure~\ref{fig:meanPt}, the centrality dependence of 
$\langle p_{T} \rangle$ for identified charged hadrons are shown
together with the 130 GeV data points~\cite{PPG006}. 
In both the 200 GeV and 130 GeV data, $\langle p_{T} \rangle$ for
all particle species increases from the most peripheral to the most 
central events and also increases with particle mass. The dependence of 
$\langle p_{T} \rangle$ on particle mass suggests the existence of 
a collective hydrodynamical expansion. The dependence on the number 
of participant nucleons ($N_{part}$) calculated using a Glauber 
model~\cite{Glauber} may be due to an increasing 
radial expansion from peripheral to central events.

\begin{figure}[tb]
\begin{minipage}[b]{0.6\linewidth}
\begin{center}
\includegraphics[scale=0.35]{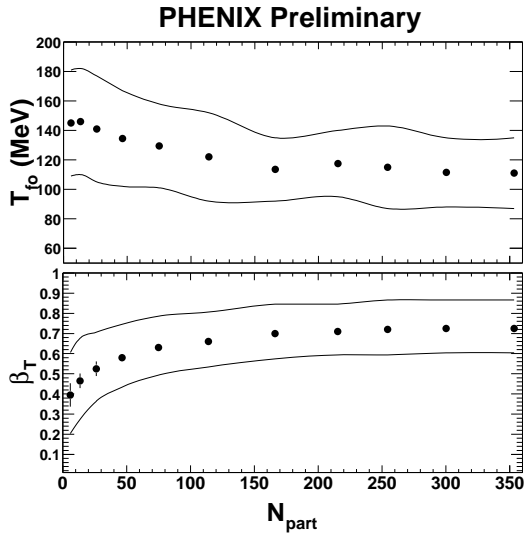}
\end{center}
\vspace{-0.4in}
\end{minipage}
\hfill
\parbox[b]{0.35\textwidth}{\sloppy
\caption{Freeze-out temperature (top) and transverse
flow velocity (bottom), as a function of $N_{part}$, extracted
from the hydrodynamical collective flow model fit to the single 
particle spectra at $\sqrt{s_{NN}}$ = 200 GeV in Au+Au.}
\protect\label{fig:HydroFit}}
\end{figure}

Using these $p_{T}$ distributions, one can characterize the collision 
system with a small set of hydrodynamic quantities based on the expanding 
source model~\cite{fit_modle}: transverse flow velocity ($\beta_{T}$) and 
freeze-out temperature ($T_{fo}$). By using this model, which 
assumes boost invariance and a linear velocity profile, a simultaneous 
fit has been performed for the measured spectra in 200 GeV Au+Au in eleven 
different centrality selections~\cite{Jane}. For the 0--5\% central events, 
an expansion flow velocity on the surface, $\beta_T$, of
0.7 $\pm$ 0.2 (syst.) and a common freeze-out temperature, $T_{fo}$, of 
110 $\pm$ 23 (syst.) MeV is extracted. The fit results of all particles 
within each event centrality are shown in Figure~\ref{fig:HydroFit}. The 
top panel is for $T_{fo}$ and the bottom panel is for $\beta_T$, both plotted 
as a function of $N_{part}$. Within the systematic uncertainties, the 
expansion parameters $T_{fo}$ and $\beta_T$ decrease and increase, 
respectively, with the number of participants, saturating at mid-centrality. 

\section{TWO-PARTICLE HBT CORRELATIONS}
Another way to extract information about the particle emitting 
source created in relativistic heavy ion collisions is with two-particle HBT
correlations. The HBT measurement is sensitive to the space-time 
evolution and duration time of the system at the freeze-out stage. 
Using the large data sample of pion pairs collected in Run II, 
we extend our HBT measurements up to 1.2 GeV/$c$ in the transverse 
momentum of the pairs ($k_{T}$) for pions~\cite{Enoki}. In this analysis 
the Bertsch-Pratt parameterization is employed in a longitudinal 
co-moving system (LCMS), where the three-dimensional Gaussian 
radius parameters are $R_{side}$, $R_{out}$ and $R_{long}$. 
These radius parameters are studied as a function of $k_{T}$ and 
centrality. We use 50 million minimum bias events with charged pions 
identified using the time-of-flight measurement in the EMC. A full Coulomb 
correction is applied assuming a Gaussian source and no pairs coming 
from resonance decays. 

\begin{figure}[h]
\begin{center}
\includegraphics[width=14cm]{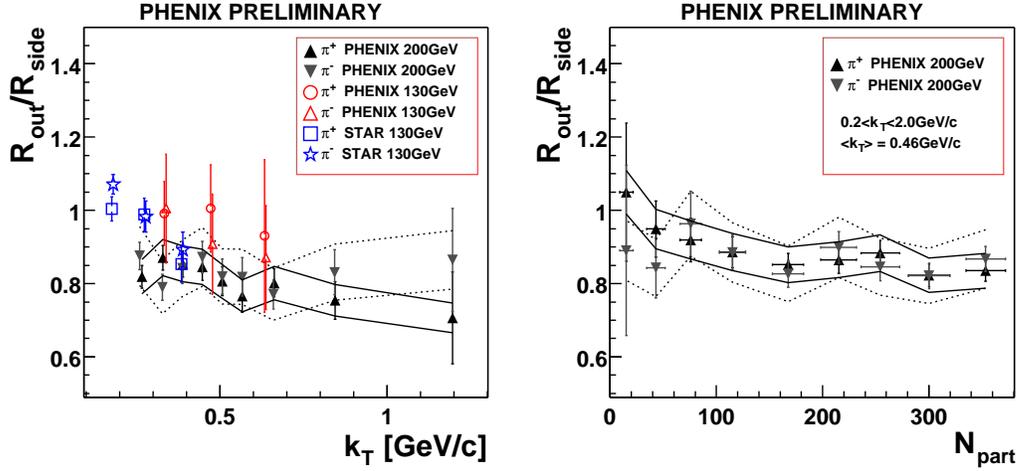}
\caption{The ratio $R_{out}/R_{side}$ for pions as a function of 
$k_T$ (left) and $N_{part}$ (right) with statistical error bars (200 GeV
data only) and a systematic error band. In the left figure, results from 
PHENIX and STAR at 130 GeV are also shown~\cite{PPG010,STAR_HBT}.}
\label{fig:HBT2}
\end{center}
\end{figure}

As shown in reference~\cite{Enoki}, we have observed a clear 
$k_{T}$ dependence for all Bertsch-Pratt radius parameters for pions in 200 
GeV Au+Au collisions. This $k_{T}$ dependence of the radius parameters 
can be explained by a space-momentum correlation effect due to the 
expansion of the system~\cite{HBT_expansion}. For the study of the duration
time of the system, the ratio $R_{out}/R_{side}$ can be used. In 
Figure~\ref{fig:HBT2}, the ratio $R_{out}/R_{side}$ as a function of 
$k_{T}$ (left) and $N_{part}$ for pion pairs is shown. The ratio does 
not change as a function of both $k_{T}$ and centrality within the 
experimental uncertainties. In contrast to the success of the hydrodynamic 
model for $p_{T}$ distributions in Section 2 of this paper, the HBT radii
from the hydrodynamic calculation agree only qualitatively with the
data with significant quantitative discrepancies~\cite{Heinz}. 
At this time, no $\chi^{2}$ minimum could be found for
the simultaneous hydrodynamic fit of the particle spectra and the 
$k_{T}$ dependence of HBT radius parameters for pions~\cite{Jane}.

\section{DEUTERON AND ANTI-DEUTERON SPECTRUM}
Deuteron and anti-deuteron measurements provide complementary
information to the HBT measurements because the coalescence coefficient 
$B_{2}$~\cite{Coal_modle} is considered to be proportional to the inverse 
of the volume of the system, and the dynamical evolution of the system
can be studied with the $p_{T}$ dependence of $B_{2}$, similar to 
two-particle HBT measurements. 

\begin{figure}[t]
\begin{minipage}[t]{80mm}
\includegraphics[width=8cm]{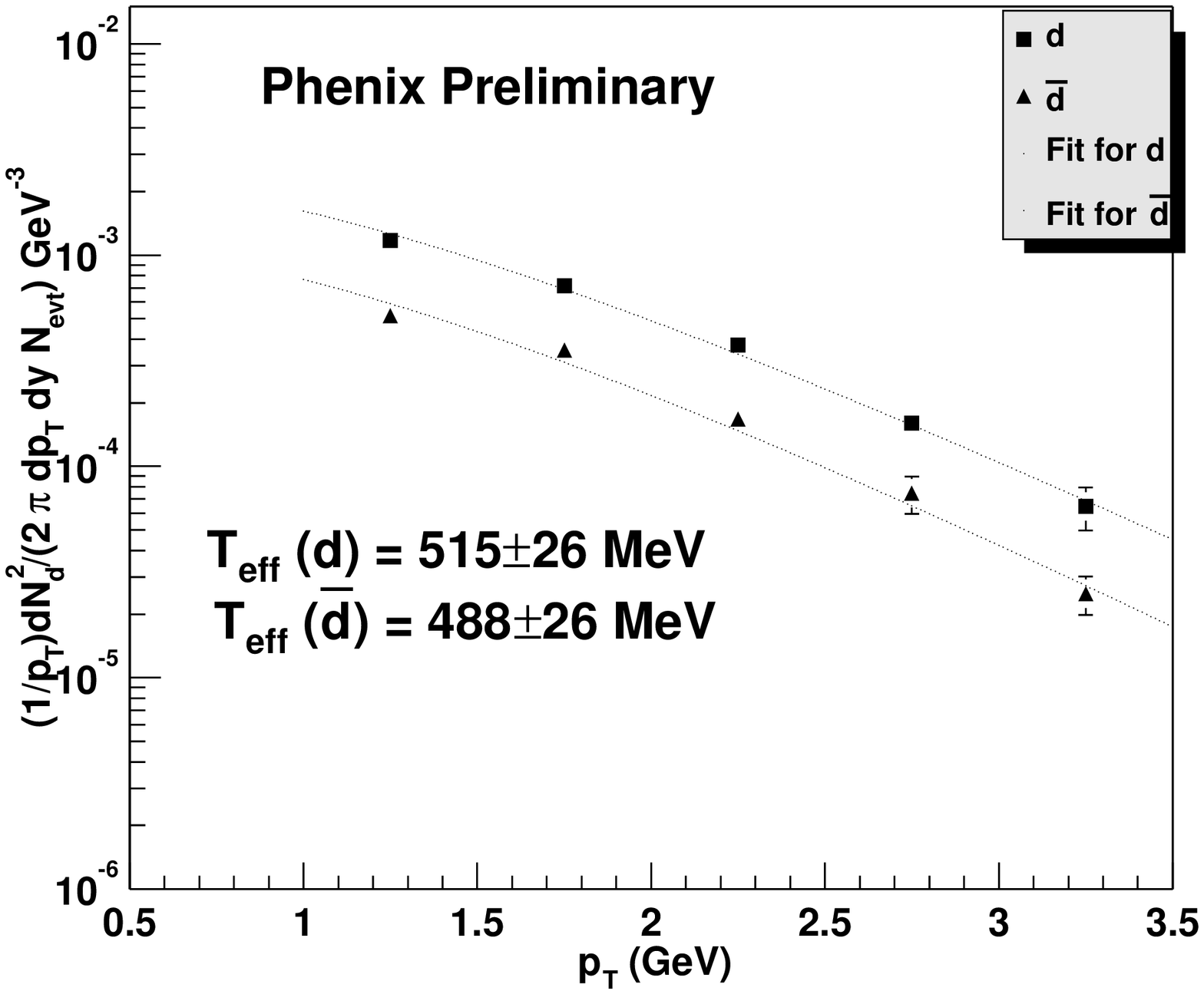}
\caption{$p_{T}$ distributions for deuterons (squares) 
and anti-deuterons (triangles). The error bars are statistical only. 
The lines are fits using an $m_T$ exponential function.}
\label{fig:Deuteron1}
\end{minipage}
\hspace{\fill}
\begin{minipage}[t]{75mm}
\includegraphics[width=8cm]{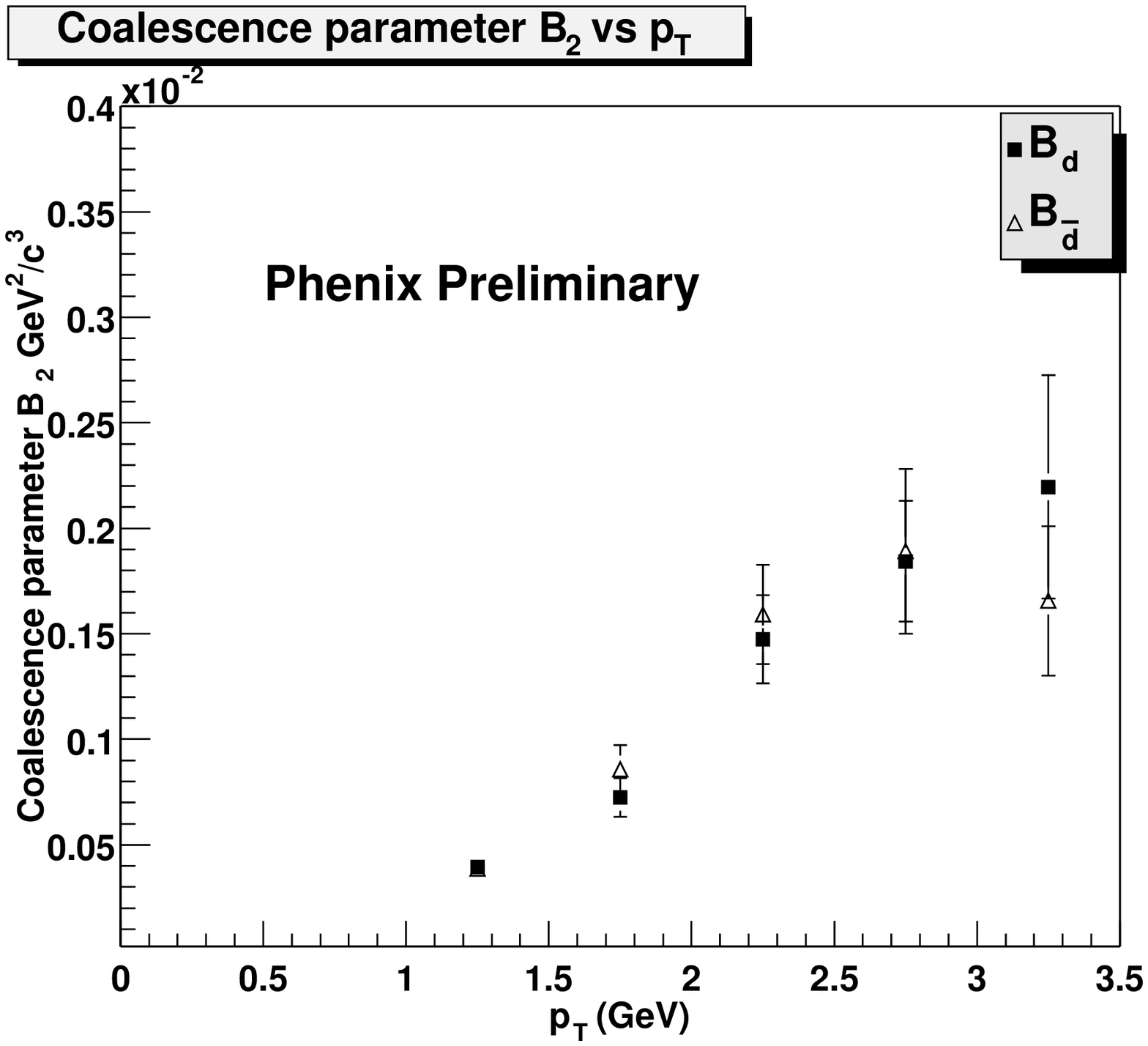}
\caption{$p_{T}$ dependence of the coalescence parameter, $B_{2}$, 
for deuterons (squares) and anti-deuterons (triangles) at 
mid-rapidity.}
\label{fig:Deuteron2}
\end{minipage}
\end{figure}

Figure~\ref{fig:Deuteron1} shows the $p_{T}$ distributions of $d$ 
(squares) and $\overline{d}$ (triangles) for minimum bias events in
200 GeV Au+Au collisions, identified with the high resolution TOF detector. 
The fitted lines using an $m_{T}$ exponential function are also shown. 
We obtain an effective temperature of 515 $\pm$ 26 (stat.) MeV for 
deuterons and 488 $\pm$ 26 (stat.) MeV for $\overline{d}$. 
Figure~\ref{fig:Deuteron2} shows the transverse momentum dependence of 
$B_{2}$ for deuterons (square) and anti-deuterons (triangle) 
at mid-rapidity. $B_{2}$ increases linearly with $p_{T}$ up to 3.5 GeV/$c$.
The $p_{T}$ dependence of $B_{2}$ is consistent with the expanding
source picture. A similar behavior has been observed in the HBT
radius parameters for pions~\cite{Enoki}.  

\section{PARTICLE RATIOS}

\begin{figure}[h]
\begin{center}
\includegraphics[width=14cm]{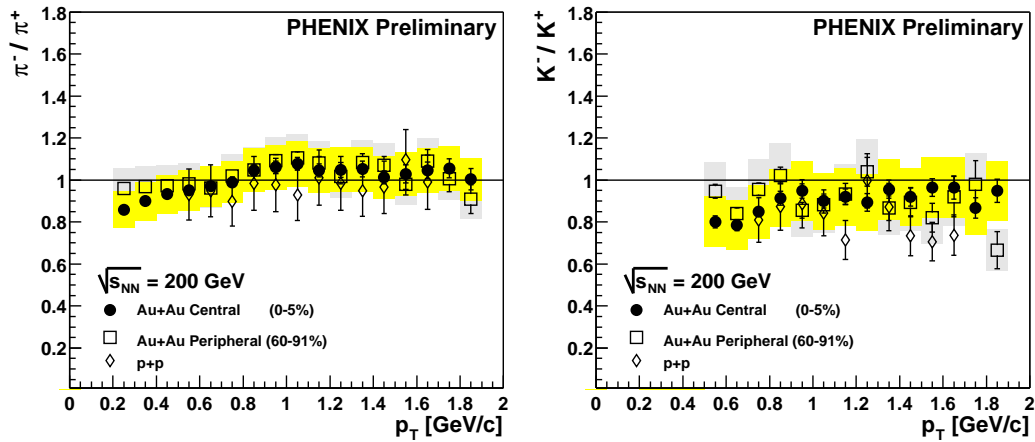}
\caption{Particle ratios of $\pi^{-}/\pi^{+}$ (left) and
$K^{-}/K^{+}$ (right) as a function of $p_{T}$. The 
data points are for the most central (filled circles) and
the most peripheral (open squares) in Au+Au and 
proton-proton (open diamonds) collisions at $\sqrt{s_{NN}}$ = 200 GeV.}
\label{fig:pi_k_ratio}
\end{center}
\end{figure}

\begin{figure}
\begin{center}
\includegraphics[width=13cm]{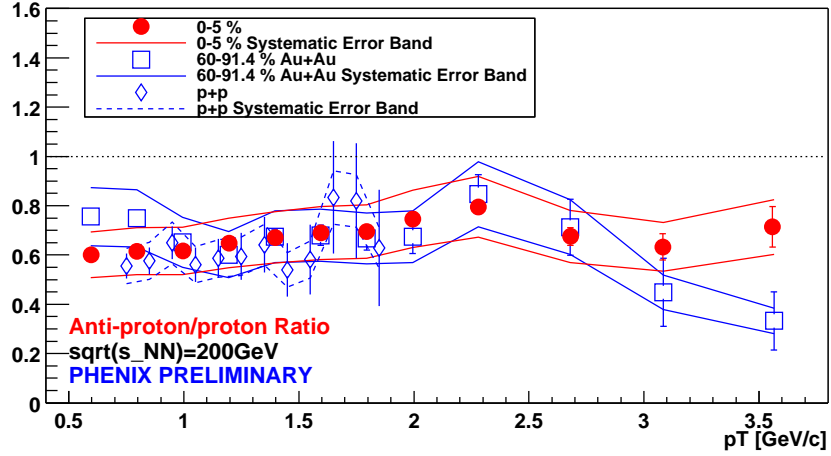}
\caption{$\overline{p}/p$ ratios as a function of $p_T$ for central 
0--5\% (filled circles), peripheral 60--91.4\% (open squares) 
Au+Au and proton-proton (open diamonds) collisions at $\sqrt{s_{NN}}$ 
= 200 GeV. The error bars indicate the statistical errors and the lines 
indicate the systematic errors for each data point.}
\label{fig:pbar_p_ratio}
\end{center}
\end{figure}

Particle ratios provide us with information about the chemical 
properties of the collision system. In Figure~\ref{fig:pi_k_ratio}, 
the particle ratios of $\pi^{-}/\pi^{+}$ (left) and $K^{-}/K^{+}$ 
(right) are shown as a function of $p_T$ for central (0--5\%) and
peripheral (60--91.4\%) Au+Au and proton-proton collisions 
at $\sqrt{s_{NN}}$ = 200 GeV. A similar plot for the $\overline{p}/p$ 
ratio is shown in Figure~\ref{fig:pbar_p_ratio}~\cite{Takao}. 
Regardless of the particle species, centrality, or collision system, 
the particle ratios are almost flat as a function of $p_{T}$ within the 
systematic errors and measured $p_T$ range, with the exception of the 
$\overline{p}/p$ ratio for peripheral events, which seems to decrease 
in the high $p_{T}$ region. The integrated ratios over the measured $p_T$ 
range in the most central events in 200 GeV Au+Au collisions are:
1.02 $\pm$ 0.02 (stat.) $\pm$ 0.1 (sys.) for $\pi^{-}/\pi^{+}$, 
0.92 $\pm$ 0.03 (stat.) $\pm$ 0.1 (sys.) for $K^{-}/K^{+}$, and
0.70 $\pm$ 0.04 (stat.) $\pm$ 0.1 (sys.) for $\overline{p}/p$. 
Based on a simple statistical thermal model~\cite{thermal}, a
baryon chemical potential, $\mu_{B}$, of $\sim$ 30 MeV for 200 GeV Au+Au 
central collisions is estimated from the $K^{-}/K^{+}$ and $\overline{p}/p$ 
ratios.

\begin{figure}
\begin{center}
\includegraphics[width=16cm]{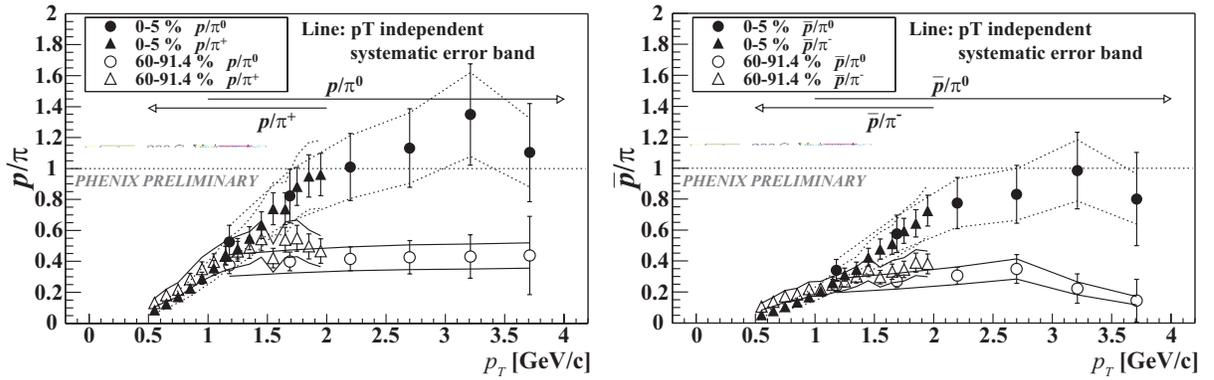}
\caption{ $p/\pi$ (left) and $\overline{p}/\pi$ (right) ratios 
as a function of $p_T$ for the 0--5 \% and 60--91.4\% centrality selections. 
Lines along the data points show the $p_T$-independent systematic error bands.
The error bars show the statistical and $p_T$-dependent systematic errors, 
summed in quadrature.}
\label{fig:ppi_ratio}
\end{center}
\end{figure}

Particle composition at high $p_T$ is also interesting in order to understand 
baryon production and transport, system evolution, and the interplay 
between soft processes and jet quenching in hard processes. In 
Figure~\ref{fig:ppi_ratio},
the $p/\pi$ and $\overline{p}/\pi$ ratios are shown. For the denominator 
of the ratio $p/\pi$ and $\overline{p}/\pi$, we use
two independent measurements with different subsystems: (1) $p_T$ 
spectra for $\pi^{\pm}$ up to 2 GeV/$c$ identified using the TOF, 2) neutral 
pions results from 1 GeV/$c$ to 3.8 GeV/$c$ measured using the EMC. 
Both the $p/\pi$ and $\overline{p}/\pi$ ratios have a clear centrality 
dependence. The data shows that these ratios in central collisions reach
unity at a $p_{T}$ of 2$\sim$3 GeV/$c$, while they saturate at around 
$p_{T}$ of 0.3 -- 0.4 in peripheral collisions. This observed behavior in 
central events may be attributed to the composition of two effects including 
a larger flow effect for protons (and $\overline{p}$) compared to pions, and 
a pion suppression effect at high $p_{T}$~\cite{PPG003}.

\section{ELLIPTIC FLOW FOR IDENTIFIED HADRONS}
In non-central collisions, the initial overlap region of two nuclei
is elliptically deformed in the transverse plane, resulting in 
anisotropic pressure gradients. These cause a more rapid expansion
into the reaction plane than perpendicular to it, resulting in an 
anisotropy of the final $p_{T}$ distributions called elliptic
flow~\cite{Heinz}. Since the event anisotropy is considered to be
developed at the early stage of collisions, the study of elliptic
flow provides a good tool to investigate the possible formation of
a QGP state. 

\begin{figure}[h]
\begin{center}
\includegraphics[width=14cm]{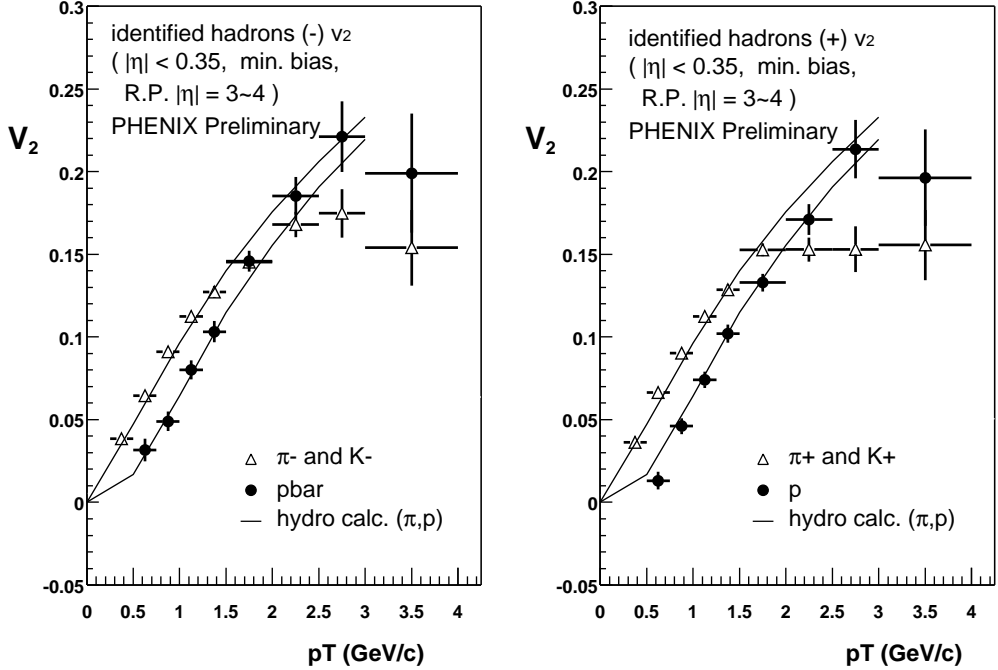}
\caption{Transverse momentum dependence of $v_2$ with respect 
to the reaction plane for ($\pi^{-}+K^{-}$) and $\overline{p}$ 
(left), and for ($\pi^{+}+K^{+}$) and $p$ (right).
The solid lines represent the hydrodynamic calculation~\cite{houv01} for
pions (upper curve) and protons (lower curve).} 
\label{fig:Flow}
\end{center}
\end{figure}

Elliptic flow is quantified by the second harmonic coefficient
($v_{2}$) of a Fourier expansion in the azimuthal distributions of 
the measured spectrum with respect to the reaction plane. 
We have measured the $v_2$ parameter for identified particles with respect to 
the reaction plane, which is defined in the beam counters 
($|\eta| = 3 \sim 4$) in 200 GeV Au+Au collisions. The details of the analysis
method and results are shown in~\cite{Esumi}. Figure~\ref{fig:Flow} shows 
the $p_{T}$ dependence of the $v_2$ parameter for identified particles with 
respect to the reaction plane for minimum bias events. The solid circles are 
for protons and $\overline{p}$, and the open triangles are for the combined
results for pions and kaons. The left panel shows negatively charged 
particles and the right panel shows positively charged particles. The solid 
lines are the results of a hydrodynamic calculation including a first order 
phase transition with a freeze-out temperature of 120 MeV~\cite{houv01}. 
The data shows that at lower $p_{T}$ ($<$ 2 GeV/$c$), the light mesons have a 
larger $v_2$ parameter compared to protons and $\overline{p}$. The model 
calculation agrees very well with the data for all particles in this
$p_T$ range. However, the data seems to deviate from the hydrodynamic 
calculations above 2 GeV/$c$ for mesons and baryons. In addition, the model 
cannot reproduce the opposite mass dependence of the $v_2$ parameter, which 
is observed above 2 GeV/$c$. 

\section{SUMMARY}
We present results on identified hadrons in Au+Au collisions at 
$\sqrt{s_{NN}}$ = 200 GeV at mid-rapidity over different centrality 
selections from the PHENIX experiment. The transverse momentum distributions 
for $\pi^{\pm}$, $K^{\pm}$, $p$, $\overline{p}$, $d$, and $\overline{d}$ 
are measured and we observe a mass dependence of the slopes and
shapes of the identified charged spectra in central events. 
However, they are almost parallel to each other in the most peripheral events. 
Also, the mean transverse momentum for all particle species increases from 
peripheral to central events, and with particle mass. In two-particle 
HBT correlations for pions, there is a clear $k_T$ dependence of all radius 
parameters. These results on the single particle spectra and two-particle HBT 
correlations are qualitatively consistent with a hydrodynamic collective 
expanding source picture, although there are quantitative discrepancies 
between the model and the HBT results. To characterize the collision system 
created in Au+Au collisions at RHIC energy, the flow velocity and freeze-out 
temperature have been extracted within the framework of a hydrodynamic 
expanding source model. 

We also present results on the particle ratios $\pi^{-}/\pi^{+}$, 
$K^{-}/K^{+}$, and $\overline{p}/p$ as a function of $p_T$ for different 
centralities in Au+Au and proton-proton collisions. It is found that the 
ratios are almost constant regardless of particle species, centrality, 
and collision system. The ratios $p/\pi$ and $\overline{p}/\pi$ up to 3.8 
GeV/$c$ are also measured by using the combined results for charged and 
neutral pions. In central collisions, the ratio reaches unity at a $p_{T}$ 
of 2$\sim$ 3 GeV/$c$ and saturates at a $p_{T}$ around 0.3 -- 0.4 GeV/$c$ 
in peripheral collisions. 

Finally, we present elliptic flow measurements with respect to the reaction 
plane for identified particles. The data show that at low $p_{T}$ 
($<$ 2 GeV/$c$) the light mesons have a larger $v_2$ parameter compared to 
protons and $\overline{p}$. The hydrodynamic model calculation agrees very 
well with the data for all particles up to 2 GeV/$c$. However, the data 
seems to deviate from the hydrodynamic model above 2 GeV/$c$ for both mesons 
and baryons. The model cannot reproduce the opposite mass dependence of the 
$v_2$ parameter above 2 GeV/$c$.

\end{document}